\documentclass[epj]{svjour}
\usepackage{graphics}
\usepackage{amsmath}
\usepackage{amssymb}
\begin{document}
\title{Neutron star structure from QCD}
\author{Eduardo S.~Fraga\inst{1} \and Aleksi Kurkela\inst{2,3} \and Aleksi Vuorinen\inst{4}
}                     
%
%
\institute{Instituto de F\'\i sica, Universidade Federal do Rio de Janeiro,
Caixa Postal 68528, 21941-972, Rio de Janeiro, RJ, Brazil \and Theory Division, PH-TH, Case C01600, CERN, CH-1211 Geneva 23, Switzerland \and Faculty of Science Technology, University of Stavanger, 4036 Stavanger, Norway \and Helsinki Institute of Physics and Department of Physics, P.O.~Box 64, FI-00014 University of Helsinki, Finland}
\date{Received: date / Revised version: date}
%
\abstract{
In this review article, we argue that our current understanding of the thermodynamic properties of cold QCD matter, originating from first principles calculations at high and low densities, can be used to efficiently constrain the macroscopic properties of neutron stars. In particular, we demonstrate that combining state-of-the-art results from Chiral Effective Theory and perturbative QCD with the current bounds on neutron star masses, the Equation of State of neutron star matter can be obtained to an accuracy better than 30\%  at all densities.
} 
\maketitle
%
\section{Introduction}
\label{intro}

Neutron stars represent a rare example of systems whose macroscopic structure is determined via a subtle interplay between the physics of vastly different length and energy scales, namely those of the strong nuclear force and gravity. It is exactly the matching of these two scales that makes  describing their properties at the same time so challenging and so rewarding; in essence, neutron stars function as natural macroscopic laboratories of nuclear physics. The task of figuring out the composition of the stars reduces to solving the well-known Tolman-Oppen\-heimer-Volkov (TOV) equations \cite{Glendenning:2000}, which need as input the equation of state (EoS) of cold and dense strongly interacting matter. This function, on the other hand, is available from the underlying microscopic theory of the strong interactions, Quantum Chromodynamics (QCD).

The problem in the above, conceptually rather straightforward story is of course the complexity of QCD. In particular, the theory has so far avoided a nonperturbative first principles solution at nonzero baryon density due to the well-known Sign Problem of lattice QCD \cite{deForcrand:2010ys}. In the absence of other nonperturbative tools, the remaining first principles options are limited to various limits: at low density --- typically below the nuclear saturation density --- the power counting of Chiral Perturbation Theory allows the rigorous construction of an effective description, Chiral Effective Theory (CET)  \cite{Epelbaum:2008ga}, to account for the two- and higher-body forces between nuclei and the construction of a reliable EoS of nuclear matter. In the opposite limit, of asymptotically high density, the asymptotic freedom of QCD guarantees that the interactions between quarks become weak and that a perturbative description of the bulk thermodynamics, i.e.~perturbative QCD (pQCD), becomes valid  \cite{Kraemmer:2003gd}. Between these limits, the range of computational tools is at the moment very limited.

In the review article at hand, our goal is two-fold. First, we want to provide a review of current perturbative calculations of the EoS of cold quark matter, and in particular to discuss the potential future developments in this field. At the same time, we also wish to illustrate how a combination of the current state-of-the-art pQCD results together with the CET EoS of nuclear matter can be used to place very stringent constraints on the behavior of the neutron star matter EoS, and thereby also on the structure of the stars. In the latter process, we follow the approach of \cite{Kurkela:2014vha}, where interpolating polytropes were used to parameterize (our ignorance of) the EoS in the region between the CET and pQCD. The criterion that this article used for switching from the low- and high-density EoSs to the polytropes was that the relative errors of the two approaches are $\pm 24\%$. Somewhat remarkably, it then follows that the simple requirements of thermodynamic stability, subluminality and the ability to support a two solar mass star \cite{demorest,antoniadis} are enough to constrain the EoS to a $\pm 30\%$ accuracy everywhere. The result of this procedure is illustrated in the schematic fig.~1.

\begin{figure*}[t]
\vspace{1.5cm}
\hspace{3.0cm}\resizebox{0.6\textwidth}{!}{%
\includegraphics{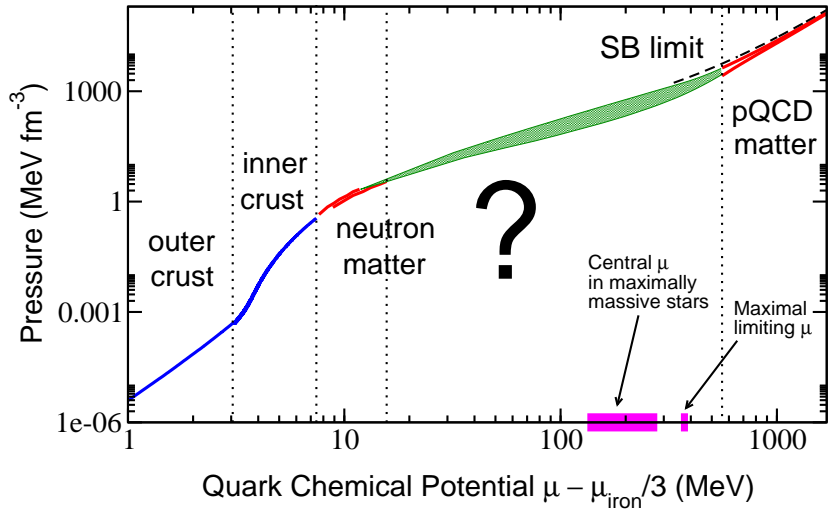}}
\caption{Known limits of the stellar EoS on a logarithmic scale. On the horizontal axis we have the quark chemical potential (with an offset so that the variable acquires the value $0$ for pressureless nuclear matter), and on the vertical axis the pressure. The band in the region around the question mark corresponds to the interpolating polytropic EoS used in \cite{Kurkela:2014vha}.}
\label{cartoon}  
\end{figure*}

An important implication of the above exercise is that the behavior of the perturbative EoS of quark matter contributes to that of neutron star matter irrespective of whether deconfined quark matter is present inside neutron stars or not. The existence of such quark matter cores is naturally a very important question by itself (see e.g.~ref.~\cite{Buballa:2014jta} for a recent review), but also considerably more challenging to approach in a model independent way than the behavior of the EoS itself. 

The structure of our article is as follows. First, in section 2 we review the current status and prospects of pQCD both at high temperatures and high densities, concentrating on the lessons to be learned from studies of high-$T$ Quark Gluon Plasma (QGP). Next, section 3 is devoted to explaining the matching and interpolation procedure of \cite{Kurkela:2014vha}, while in section 4 we review the implications of these studies on neutron star structure. Section 5 finally contains concluding remarks as well as our view of where future efforts in the field should be directed, if we want to decrease the current uncertainty of the EoS of neutron star matter.

\section{Equilibrium thermodynamics of cold quark matter}
\label{sec:1}
The thermodynamic properties of deconfined quark matter has been a topic of active research for decades. In the regime of high temperatures, the motivation stems from ultrarelativistic heavy ion physics and the early universe, while at lower temperatures and high densities the primary motivator has been the desire to understand the composition and properties of compact stars. While at high temperatures the leading source of information is by now unequivocally lattice QCD, at high densities its use is prevented by the famous Sign Problem, leaving the problem to be tackled by a combination of phenomenological models and perturbative approaches, as discussed above. 

In this section, we review the current status of research on the bulk thermodynamics of quark matter, and in particular explain the prospects and limitations of first principles weak coupling calculations as a means of determining the EoS of cold and dense deconfined matter. Though historically important for the development of the field, we leave the topic of non-first-principles model calculations aside in our presentation; for a classic review with plenty of references to relevant papers, see \cite{Buballa:2003qv}. The section is structured such that we first review the status of perturbation theory at high temperatures, paying attention to the agreement of the results with lattice simulations. After this, we take a look at the zero-temperature limit, and finally discuss the interpolation of perturbative results between these limiting cases as well as briefly comment on the prospects of future developments.

\subsection{Lessons from high temperatures}
\label{sec:2}

As the only nonperturbative first principles tool available, lattice QCD has established itself as the method of choice for the evaluation of thermodynamic quantities whenever numerical Monte-Carlo simulations are feasible. At vanishing baryon density, the efforts of several independent groups have indeed led to pinning down both the EoS, the (pseudo-)critical temperature of the deconfinement transition and various other quantities to a very good accuracy (see e.g.~\cite{Ding:2015ona,Bazavov:2014pvz,Borsanyi:2015waa} for recent results), and by now there is impressive agreement on all relevant observables. Proceeding away from the $\mu_B=0$ axis, the complex-valuedness of the lattice action, however, complicates things significantly, and it is only for rather small values of $\mu_B/T$ that methods such as Taylor expanding physical observables around $\mu_B=0$ \cite{Ejiri:2003dc} or statistical reweighting \cite{Fodor:2001pe} allow one to accurately estimate their behavior. At the same time, a different limit that has historically been problematic for lattice methods, namely very high temperatures, $T\gg T_c$, is by now quite well under control \cite{Borsanyi:2012ve}.

\begin{figure*}[t]
\vspace{0.8cm}
\hspace{0.5cm}\resizebox{0.45\textwidth}{!}{%
\includegraphics{fig1.eps}}
\hspace{0.1cm}
\resizebox{0.46\textwidth}{!}{%
\includegraphics{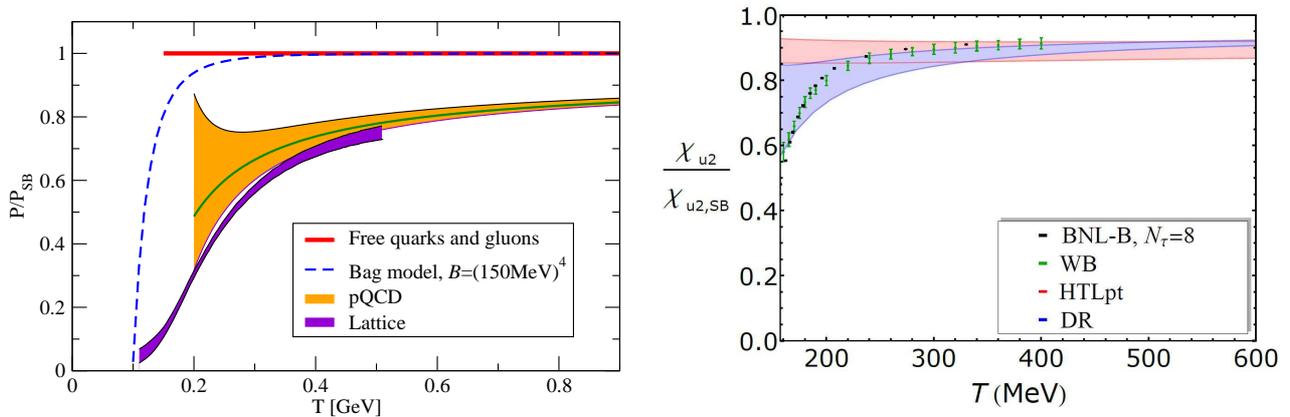}}
\caption{The $\mu_B=0$ pressure and the second order light quark number susceptibility as functions of temperature, both normalized to their respective free theory values. The lattice data are from the Wuppertal-Budapest (left figure and WB in the right figure) \cite{wuppertalcharge} and BNL-Bielefeld \cite{biele1} collaborations, while the yellow perturbative band in the left figure and the two perturbative bands of the right figure are from \cite{Laine:2006cp} and \cite{Mogliacci:2013mca}, respectively.}
\label{fig1}       
\end{figure*}

A different first principles method, with properties largely complementary to lattice QCD, is again perturbation theory, or more generally weak coupling expansions. Here, one proceeds to expand the functional integral corresponding to the partition function in a generalized power series in the coupling constant $g$, relying on the fact that asymptotic freedom guarantees that this is a well-defined procedure at least at sufficiently high energy densities. Just like with lattice simulations, there is a long history of thermal perturbation theory, dating all the way back to the late 1970's \cite{Shuryak:1977ut,Freedman:1976ub}. At present, the EoS of the QGP is known up to ${\mathcal O}(g^6\ln g)$ at high temperatures and at most moderate chemical potentials $\mu_B\leq 10T$ \cite{Kajantie:2002wa,Vuorinen:2003fs,Vuorinen:2002ue} (see also related work in Hard Thermal Loop perturbation theory \cite{Andersen:2011sf,Haque:2014rua}) and to order $g^4$ at $T=0$, including non-zero quark masses \cite{Kurkela:2009gj}. Between these limits, there exists a three-loop (i.e.~${\mathcal O}(g^4)$) result, which, however, relies on a rather heavy numerics and has only been worked out for one special case, namely a system of three massless quarks in beta equilibrium \cite{Ipp:2006ij}. 

In the region of the QCD phase diagram, where lattice QCD is applicable, a direct comparison of the predictions of lattice simulations and perturbative calculations for the EoS and quark number susceptibilities (QNSs) shows remarkably good agreement from temperatures of order $3T_c$ onwards. This is demonstrated in figure \ref{fig1}, where we display the $\mu_B=0$ pressure as well as the second order diagonal QNS as predicted by lattice QCD and resummed perturbation theory. The resummation applied in the perturbative results of fig.~\ref{fig1} (with the exception of the HTLpt band on the right) is motivated by the dimensionally reduced effective theory EQCD \cite{Kajantie:1995dw,Braaten:1995cm} that can be used to express the contribution of the soft momentum scales $gT$ and $g^2T$ --- respectively corresponding to the electro- and magnetostatic screening masses --- to the EoS. At the same time, the prediction of the MIT bag model with a commonly used bag constant $B=150$MeV, displayed for the pressure in fig.~\ref{fig1} (left), is seen to lead to a wildly differing prediction that in particular approaches the free theory limit in a rapid power-law fashion, in stark contrast with the logarithmic approach of the perturbation theory result.

An important feature of the weak coupling expansion method is that the results come with a built-in error estimator, given by their dependence on the scale parameter $\bar{\lambda}$ of the renormalization scheme in question (here the so-called modified minimal subtraction scheme). This parameter is an artifact of having had to truncate the weak coupling series after a finite number of terms, and its value is in principle completely arbitrary. As long as the perturbative expansion converges (in an asymptotic series sense), the dependence on this scale diminishes order by order, and hence it makes sense to choose some reasonable central value for it, corresponding to the dominant energy scales in the system (such as $2\pi T$ at high temperature), and gauge the uncertainty in the result by varying the parameter around this number. This is the leading source of the perturbative error bands in fig.~\ref{fig1}, and in particular explains their widening at lower temperatures, where the coupling constant of the theory grows rapidly. 

\begin{figure*}[t]
\vspace{1.29cm}
\hspace{3.0cm}\resizebox{0.6\textwidth}{!}{%
\includegraphics{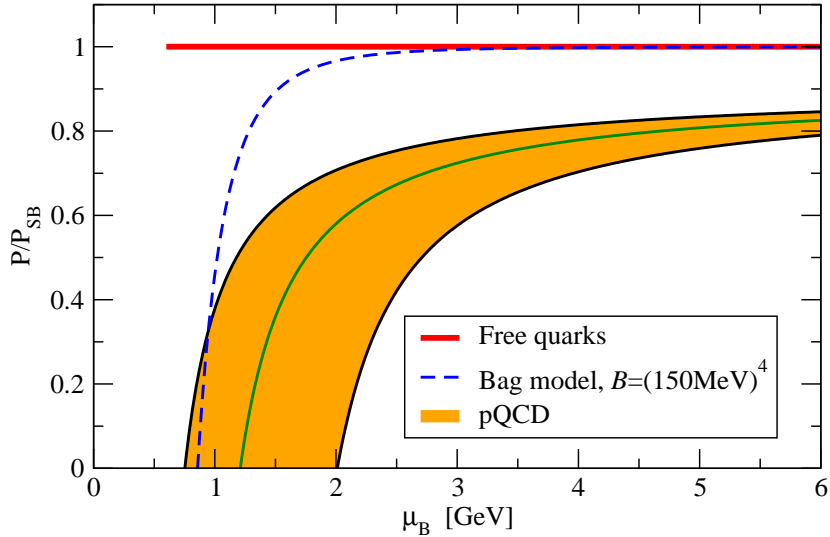}}
\caption{The pressure of $T=0$ quark matter, again normalized by the non-interacting limit. The perturbative band corresponds to the state-of-the-art three-loop calculation of \cite{Kurkela:2009gj}, including a nonzero strange quark mass.}
\label{fig2}       
\end{figure*}

\subsection{Zero temperature limit}
\label{sec:3}

Proceeding next to the zero temperature limit, relevant for neutron star physics, we no longer have the lattice QCD results available for comparison. To add to the complication, there is no longer any effective theory description available for the IR sector of the theory, and hence no natural resummation scheme that would significantly improve the convergence of the weak coupling expansion. Fortunately, some technical simplifications do occur in the exact $T=0$ limit, which allow one to efficiently use the so-called cutting rules, discussed in some length in \cite{Kurkela:2009gj}. In brief, it is possible to reduce the computation of a vacuum (bubble) diagram at zero temperature and finite chemical potential to a sum of three-dimensional numerical phase space integrals over amplitudes that are all evaluated at $\mu=0$. This presents a significant simplification to the calculations owing to the fact that results for the $\mu=0$ amplitudes can be efficiently evaluated using Integration By Parts (IBP) techniques, and are in addition abundantly available in the literature (for more discussion and references, see \cite{Kurkela:2009gj}).

The fact that the weak coupling expansion of the $T=0$ pressure has only been computed to order $g^4$ and no resummation has been carried out in it  can be seen in fig.~\ref{fig2}, where we plot the state-of-the-art perturbative result of \cite{Kurkela:2009gj} --- computed with nonzero quark masses --- and compare it to the same simple model prediction as in fig.~\ref{fig1} (left). We observe that just like in the high-temperature case, the perturbative band narrows down at high (energy) densities, but this time it is somewhat wider than in fig.~\ref{fig1} (left). Nevertheless, we observe that the error bar decreases rapidly enough for the perturbative result to have significant predictive power once  $\mu_B$ becomes of order 2.5-3 GeV.

Recently, it has been shown in \cite{Fraga:2013qra} that in beta equilibrium, the pQCD pressure can be cast in the form of a very simple pocket formula,
\begin{eqnarray}
P_{\rm{QCD}}(\mu_B,X) &= P_{\rm{SB}}(\mu_B) \left( c_1 - \frac{a(X)}{(\mu_B/{\rm GeV}) - b(X)} \right), \label{eq:pressure}\\
a(X) &= d_1 X^{-\nu_1},\quad
b(X) = d_2 X^{-\nu_2},
\end{eqnarray}
where $X$ is a parameter proportional to the renormalization scale of the theory (typically varied within $X\in[1,4]$), and we have denoted the pressure of three massless noninteracting quark flavors (at $N_c=3$) by
\begin{eqnarray}
P_{\rm SB}(\mu_B) = \frac{3}{4\pi^2}(\mu_B/3)^4.
\end{eqnarray}
The values of the constants $c_1,d_1,d_2,\nu_1,\nu_2$ are fixed by making sure that the pressure, quark number density and speed of sound squared obtained from the fit  agree with the full results of \cite{Kurkela:2009gj}, leading to
\begin{eqnarray}
c_1=0.9008 \quad & d_1= 0.5034 &\quad d_2 = 1.452 \\
\nu_1 &= 0.3553 \quad \nu_2&= 0.9101.
\end{eqnarray}
For these values, one obtains a good fit whenever $\mu_B< 2\,\rm{GeV}$, $P(\mu_B)>0$, and $X\in[1,4]$. Similar pocket formulas can be equally well derived outside the limits of beta equilibrium and charge neutrality, e.g.~for fixed lepton fraction.

The above pocket formula for the pressure allows for the derivation of a simple analytic expression for the trace anomaly $\epsilon_{\rm{QCD}} - 3P_{\rm{QCD}}$, where $\epsilon$ stands for the energy density,
\begin{equation}
\epsilon_{\rm{QCD}} - 3P_{\rm{QCD}}=\frac{\mu_B}{{\rm GeV}} P_{\rm{SB}}(\mu_B)   
\frac{a(X)}{\left[ (\mu_B/{\rm GeV}) - b(X)\right]^{2}} .
\end{equation}
For the MIT bag model, the corresponding result would be simply $4B$, i.e.~a constant. It is thus clear that a bag model description completely misses important physics, namely the degree of conformality violation in the system that is measured by the trace anomaly.

\begin{figure*}[t]
\vspace{0.4cm}
\hspace{0.5cm}\resizebox{0.45\textwidth}{!}{%
\includegraphics{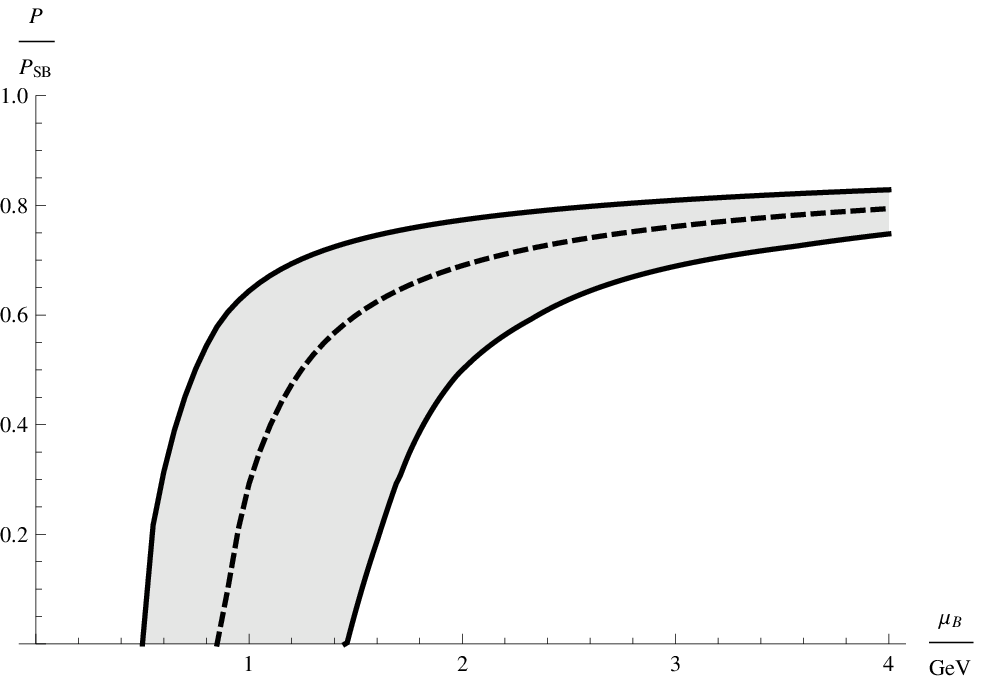}}
\hspace{0.1cm}
\resizebox{0.45\textwidth}{!}{%
\includegraphics{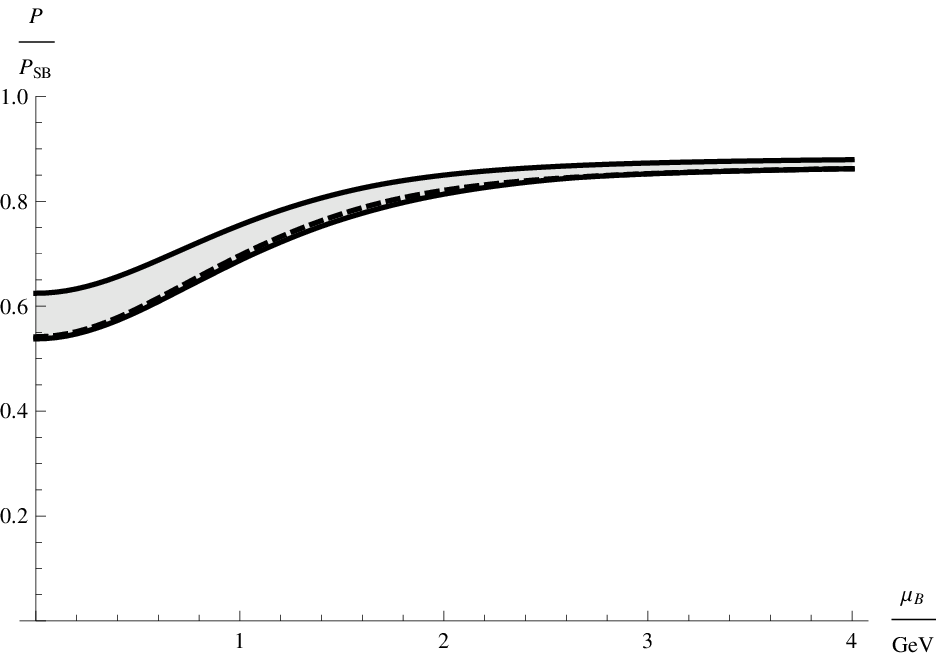}}
\caption{The pressure of a system of three quarks, the electron and the electron neutrino, evaluated as a function of the baryon chemical potential at $T=30$ and 100 MeV and a fixed lepton fraction $Y_l=0.4$. The bands are obtained through a simple interpolation between the known zero quark mass results at $T=0$ \cite{Freedman:1976ub,Vuorinen:2003fs} and high temperature \cite{Vuorinen:2003fs,Mogliacci:2013mca}.}
\label{fig3}       
\end{figure*}

\subsection{Intermediate region and future prospects}
\label{sec:4}

Somewhat counterintuitively, the technically most complicated region for thermal perturbation theory is that of high densities and small but nonzero temperatures. Here, the effective theories that allow for an efficient packaging of the IR contributions to thermodynamic quantities are not available, but neither are the computational simplifications specific to the $T=0$ limit. In the special case of three massless quarks at the same chemical potential, it was demonstrated in \cite{Ipp:2006ij} that it is possible to perform a direct all-orders resummation of certain classes of four-dimensional vacuum Feynman diagrams, and thereby obtain the EoS to the full $g^4$ order. This method is, however, technically quite cumbersome and not automatically generalizable to non-zero quark masses or other values of chemical potentials, away from the limits of charge neutrality and beta equilibrium. Thus, the technically most straightforward alternative to obtain the EoS at temperatures of order $0<T<200$ MeV is to perform a simple interpolation between the $T=0$ and high-temperature results, recalling that the latter are applicable as long as $T \gtrsim \mu_{B}/10$. The result of this procedure is shown in figure \ref{fig3}, where we display the behavior of the pressure of a system of three quark flavors, the electron and the electron neutrino at $T=30$ and 200 MeV and a fixed lepton fraction of $Y_l=0.4$. We observe a rapid decrease of the uncertainty when moving to higher temperatures, which is in part simply due to the larger energy densities there, and in part to the resummation carried out in the high-$T$ calculation. Note that the interpolation is performed between the zero quark mass $T=0$ and high-$T$ results due to the fact that the latter only exists in this limit.

As will be demonstrated in the following two sections, the most important way perturbative calculations can aid studies of the neutron star structure in the near future is by reducing the current uncertainties in the EoS of quark matter at $T=0$. To achieve this goal, the most crucial step would undoubtedly be the determination of the next orders in the weak coupling expansion: First the term of order $g^6\ln\,g$, and later the full ${\mathcal O}(g^6)$ four-loop result. While progress of this sort will not aid in building a physical resummation scheme for the EoS, there is reason to expect that it will lead to a considerable reduction of the renormalization scale dependence of the pressure. This is due to the fact that the four-loop order is the first one where an optimization of the midpoint value of the scale parameter becomes possible through schemes such as the Principle of Minimal Sensitivity (PMS) or the Fastest Apparent Convergence (FAC). Needless to say, such a four-loop computation is, however, a formidable challenge, and one that will take a considerable amount of manpower and time to tackle.

%

%
%

\begin{thebibliography}{}

\bibitem{Glendenning:2000}
  N.~K.~Glendenning,
  {\it Compact Stars -- Nuclear Physics, Particle Physics and General Relativity,}
  (Springer, 2000).
  
\bibitem{deForcrand:2010ys} 
  P.~de Forcrand,
  PoS LAT {\bf 2009}, 010 (2009).

  
\bibitem{Epelbaum:2008ga} 
  E.~Epelbaum, H.~W.~Hammer and U.~G.~Meissner,
  Rev.\ Mod.\ Phys.\  {\bf 81}, 1773 (2009).
  
\bibitem{Kraemmer:2003gd} 
  U.~Kraemmer and A.~Rebhan,
  Rept.\ Prog.\ Phys.\  {\bf 67}, 351 (2004).
  
\bibitem{Kurkela:2014vha}
  A.~Kurkela, E.~S.~Fraga, J.~Schaffner-Bielich and A.~Vuorinen,
  Astrophys.\ J.\  {\bf 789}, 127 (2014).

\bibitem{demorest} 
P. Demorest, T. Pennucci, S. Ransom, M. Roberts, and J. Hessels, 
Nature {\bf 467}, 1081Ð1083 (2010).

\bibitem{antoniadis} 
J. Antoniadis, P. C. Freire, N. Wex, T. M. Tauris, R. S. Lynch, et al., 
Science {\bf 340}, 6131 (2013).  

\bibitem{Buballa:2014jta} 
  M.~Buballa, V.~Dexheimer, A.~Drago, E.~Fraga, P.~Haensel, I.~Mishustin, G.~Pagliara and J.~Schaffner-Bielich {\it et al.},
  J.\ Phys.\ G {\bf 41}, no. 12, 123001 (2014).
  


\bibitem{Buballa:2003qv}
  M.~Buballa,
  Phys.\ Rept.\  {\bf 407}, 205 (2005).
  
\bibitem{Ding:2015ona}
  H.~T.~Ding, F.~Karsch and S.~Mukherjee,
  arXiv:1504.05274 [hep-lat].
  
\bibitem{Bazavov:2014pvz}
  A.~Bazavov {\it et al.} [HotQCD Collaboration],
  Phys.\ Rev.\ D {\bf 90} (2014) 9,  094503
  [arXiv:1407.6387 [hep-lat]].
  
\bibitem{Borsanyi:2015waa}
  S.~Borsanyi, S.~Durr, Z.~Fodor, C.~Holbling, S.~D.~Katz, S.~Krieg, D.~Nogradi and K.~K.~Szabo {\it et al.},
  arXiv:1504.03676 [hep-lat].

\bibitem{Ejiri:2003dc}
  S.~Ejiri, C.~R.~Allton, S.~J.~Hands, O.~Kaczmarek, F.~Karsch, E.~Laermann and C.~Schmidt,
  Prog.\ Theor.\ Phys.\ Suppl.\  {\bf 153} (2004) 118
  [hep-lat/0312006].
  
  
\bibitem{Fodor:2001pe}
  Z.~Fodor and S.~D.~Katz,
  JHEP {\bf 0203} (2002) 014
  [hep-lat/0106002].
  
  
  
\bibitem{Borsanyi:2012ve}
  S.~Borsanyi, G.~Endrodi, Z.~Fodor, S.~D.~Katz and K.~K.~Szabo,
  JHEP {\bf 1207} (2012) 056
  [arXiv:1204.6184 [hep-lat]].

  
  
\bibitem{Shuryak:1977ut}
  E.~V.~Shuryak,
  Sov.\ Phys.\ JETP {\bf 47} (1978) 212
   [Zh.\ Eksp.\ Teor.\ Fiz.\  {\bf 74} (1978) 408].

\bibitem{Freedman:1976ub}
  B.~A.~Freedman and L.~D.~McLerran,
  Phys.\ Rev.\ D {\bf 16} (1977) 1169.
  
\bibitem{Kajantie:2002wa}
  K.~Kajantie, M.~Laine, K.~Rummukainen and Y.~Schroder,
  Phys.\ Rev.\ D {\bf 67} (2003) 105008
  [hep-ph/0211321].
  
\bibitem{Vuorinen:2003fs}
  A.~Vuorinen,
  Phys.\ Rev.\ D {\bf 68} (2003) 054017
  [hep-ph/0305183].
  
\bibitem{Vuorinen:2002ue}
  A.~Vuorinen,
  Phys.\ Rev.\ D {\bf 67} (2003) 074032
  [hep-ph/0212283].
  
\bibitem{Andersen:2011sf}
  J.~O.~Andersen, L.~E.~Leganger, M.~Strickland and N.~Su,
  JHEP {\bf 1108} (2011) 053
  [arXiv:1103.2528 [hep-ph]].
  
\bibitem{Haque:2014rua}
  N.~Haque, A.~Bandyopadhyay, J.~O.~Andersen, M.~G.~Mustafa, M.~Strickland and N.~Su,
  JHEP {\bf 1405} (2014) 027
  [arXiv:1402.6907 [hep-ph]].
  
  
\bibitem{Kurkela:2009gj}
  A.~Kurkela, P.~Romatschke and A.~Vuorinen,
  Phys.\ Rev.\ D {\bf 81}, 105021 (2010).


  
\bibitem{Ipp:2006ij}
  A.~Ipp, K.~Kajantie, A.~Rebhan and A.~Vuorinen,
  Phys.\ Rev.\ D {\bf 74} (2006) 045016
  [hep-ph/0604060].

\bibitem{Kajantie:1995dw}
  K.~Kajantie, M.~Laine, K.~Rummukainen and M.~E.~Shaposhnikov,
  Nucl.\ Phys.\ B {\bf 458} (1996) 90
  [hep-ph/9508379].
  
\bibitem{Braaten:1995cm}
  E.~Braaten and A.~Nieto,
  Phys.\ Rev.\ D {\bf 51} (1995) 6990
  [hep-ph/9501375].








\bibitem{wuppertalcharge}
S.~Borsanyi, Z.~Fodor, S.~D.~Katz, S.~Krieg, C.~Ratti, and K.~Szabo,
\emph{JHEP} {\bf 01} (2012) 138 [arXiv:1112.4416].


  \bibitem{biele1}
C.~Schmidt, 
\emph{J.\ Phys.\ Conf.\ Ser.} {\bf 432} (2013) 012013v [arXiv:1212.4283].

\bibitem{Laine:2006cp}
  M.~Laine and Y.~Schroder,
  Phys.\ Rev.\ D {\bf 73} (2006) 085009
  [hep-ph/0603048].

\bibitem{Mogliacci:2013mca}
  S.~Mogliacci, J.~O.~Andersen, M.~Strickland, N.~Su and A.~Vuorinen,
  JHEP {\bf 1312} (2013) 055
  [arXiv:1307.8098 [hep-ph]].

  
\bibitem{Fraga:2013qra}
  E.~S.~Fraga, A.~Kurkela and A.~Vuorinen,
  Astrophys.\ J.\  {\bf 781}, L25 (2014).
  
\bibitem{Ruester:2005fm}
  S.~B.~Ruester, M.~Hempel and J.~Schaffner-Bielich,
  Phys.\ Rev.\ C {\bf 73} (2006) 035804
  [astro-ph/0509325].
  
  
  
\bibitem{Tews:2012fj}
  I.~Tews, T.~Krüger, K.~Hebeler and A.~Schwenk,
  Phys.\ Rev.\ Lett.\  {\bf 110} (2013) 3,  032504
  [arXiv:1206.0025 [nucl-th]].
 
  
  
\bibitem{Lattimer:2010uk} 
  J.~M.~Lattimer and M.~Prakash,
  arXiv:1012.3208 [astro-ph.SR].
 
 
 
\bibitem{Hebeler:2013nza} 
 K. Hebeler, J.~M. Lattimer, C.~J. Pethick and A. Schwenk,
  Astrophys.\ J.\  {\bf 773}, 11 (2013).
 
 
\bibitem{Alford:2013aca}
  M.~G.~Alford, S.~Han and M.~Prakash,
  Phys.\ Rev.\ D {\bf 88} (2013) 8,  083013
  [arXiv:1302.4732 [astro-ph.SR]].
 
\bibitem{Alford:2015dpa}
  M.~G.~Alford, G.~F.~Burgio, S.~Han, G.~Taranto and D.~Zappalà,
  arXiv:1501.07902 [nucl-th].
 
\bibitem{Fraga:2001id} 
  E.~S.~Fraga, R.~D.~Pisarski and J.~Schaffner-Bielich,
  Phys.\ Rev.\ D {\bf 63}, 121702 (2001)
  [hep-ph/0101143].

\bibitem{Fraga:2001xc} 
  E.~S.~Fraga, R.~D.~Pisarski and J.~Schaffner-Bielich,
  Nucl.\ Phys.\ A {\bf 702}, 217 (2002)
  [nucl-th/0110077].
  
\bibitem{Alford:2004pf} 
  M.~Alford, M.~Braby, M.~W.~Paris and S.~Reddy,
  Astrophys.\ J.\  {\bf 629}, 969 (2005)
  [nucl-th/0411016].
  
\bibitem{Fraga:2004gz} 
  E.~S.~Fraga and P.~Romatschke,
  Phys.\ Rev.\ D {\bf 71}, 105014 (2005)
  [hep-ph/0412298].
  
\bibitem{Kurkela:2010yk} 
  A.~Kurkela, P.~Romatschke, A.~Vuorinen and B.~Wu,
  arXiv:1006.4062 [astro-ph.HE].
  
\end{thebibliography}
%

\section{Bridging the gap between nuclear and pQCD matter}

\begin{figure*}[t]
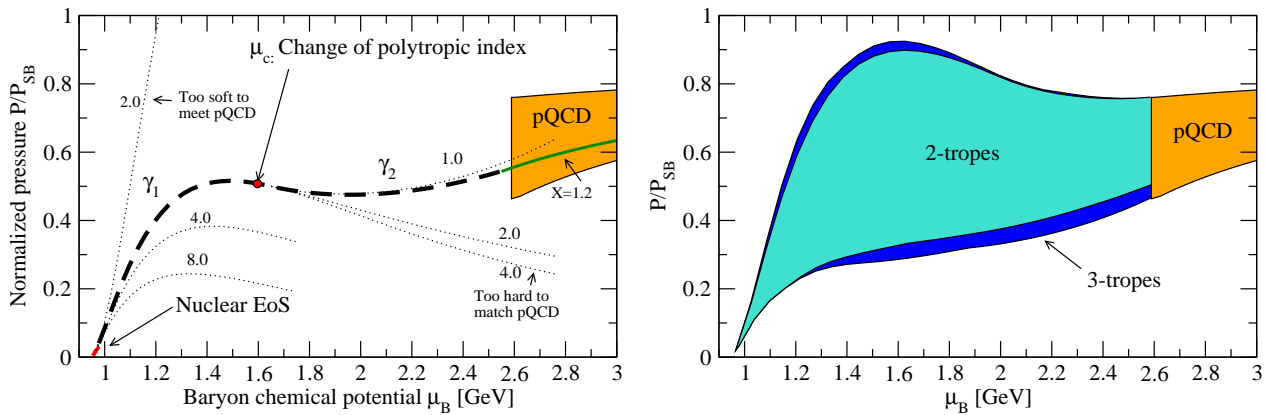

\vspace{0.4cm}
\hspace{0.5cm}\resizebox{0.45\textwidth}{!}{%
\includegraphics{eos_polytrope.eps}}
\hspace{0.1cm}
\resizebox{0.45\textwidth}{!}{%
\includegraphics{fig7.eps}}
\caption{Left: The matching of the low and high density limiting regimes using $2$-tropes. Solutions exist for $\mu_{c} \in [1.08,2.05]$, $\gamma_{1} \in [2.23,9.2]$ and $\gamma_{2} \in [1.0,1.5]$. Right: The change in the EoS cloud (without a mass constraint) when allowing for a third interpolating monotrope between the two regimes.}
\label{fig5}       
\end{figure*}

As explained in section 1, figure \ref{cartoon} summarizes the current first principles information available on the behavior of the EoS of zero temperature strongly interacting matter. At the lowest densities, the system corresponds to a lattice of nuclei of increasing neutron fraction, which continues until the neutron drip line is reached in the inner crust of the star \cite{Ruester:2005fm}. At higher chemical potentials, the matter on the other hand consists predominantly of neutrons, whose interactions gradually become more important. The uncertainties in the EoS of neutron matter rapidly grow with density such that the function is known within $\pm 24\%$ at the density of $1.1$ times the nuclear saturation density $n_0$ \cite{Tews:2012fj}. This density corresponds to $\mu_B \in [0.968,0.978]$GeV,  to be contrasted with the chemical potentials found inside stellar cores. 

The value of the chemical potential in the center of a maximally massive star is a strongly model dependent quantity. A strict upper limit can, however, be given by an elegant argument based solely on the maximally stiff EoS and the TOV equation, limiting the chemical potentials to $\mu_B\lesssim 2.1$GeV \cite{Lattimer:2010uk}; this limit is denoted in fig.~\ref{cartoon} by the label ``Maximal limiting $\mu$''. The physical EoS, however, is surely not maximally stiff, and typical values for the central chemical potential range within $\mu_B \approx [1.33,1.84]$GeV, denoted by the magenta line in fig.~\ref{cartoon}.  These chemical potentials are clearly far out of reach for the present-day nuclear physics calculations. Strong extrapolation is therefore needed in order to construct the EoS of a maximally massive star using only the low density information, which suggests it should be very useful to approach the relevant region of densities also from the high density side.

The EoS of free quarks is displayed in fig.~1 as a dashed line labeled ``SB limit'', standing for the Stefan-Boltzmann limit. This piece of information alone offers a constraint for the high-density EoS by providing the limit to which EoS must eventually asymptote. However, the constraint is rather weak because it does not include information about \emph{how} the physical EoS is supposed to approach this asymptote, and in particular at what chemical potentials the free limit becomes numerically relevant. The higher order corrections to the EoS, discussed in the previous section, however ameliorate this difficulty, allowing us to estimate the accuracy and precision of the perturbative result.

The baryon number chemical potentials where the uncertainties of the perturbative calculation become comparable to those of the low energy EoS are around $\mu_B \approx 2.6$GeV. While this value is clearly larger than the chemical potentials expected to be found inside the cores of maximal neutron stars, and even larger than the limiting chemical potential, it nevertheless gives a strong constraint for the behavior of the EoS at high densities. 

\begin{figure*}[t]
\vspace{0.0cm}
\hspace{3.0cm}\resizebox{0.6\textwidth}{!}{%
\includegraphics{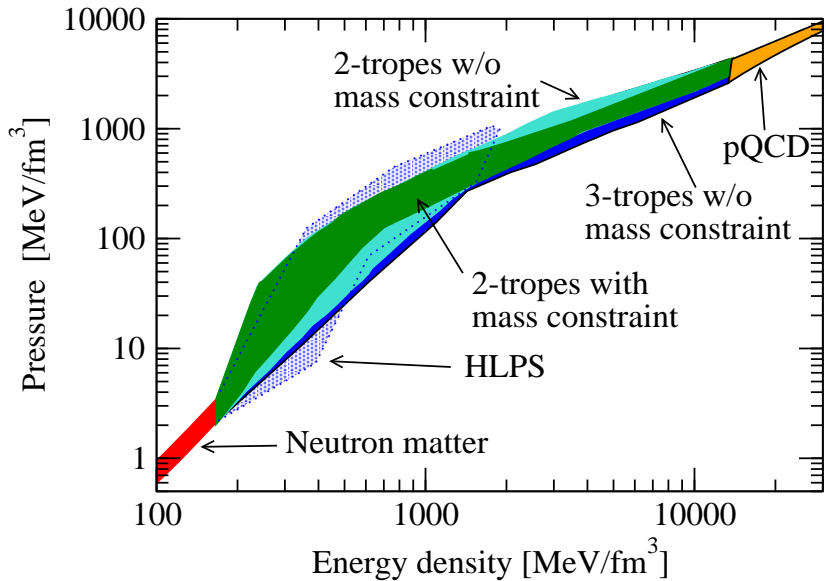}}
\caption{The equation of state $p=p(\epsilon)$, obtained from different interpolation and extrapolation schemes. The green, turquoise and blue bands correspond to our result, while the light blue `HLPS' band stands for the result of \cite{Hebeler:2013nza}.}
\label{fig6}       
\end{figure*}

To quantify the power of the additional information coming from pQCD, an interpolation between the two presently known limits was considered in \cite{Kurkela:2014vha}. In the intermediate region, the EoS of neutron star matter is parameterized using a polytropic interpolator, constructed from a set of monotropes $P_i(n)= \kappa_i n^{\gamma_i}$, with several (baryon number) density intervals described by different polytropic indices $\gamma_i$. The larger $\gamma_i$ is, the stiffer the EoS is, while for a given $\gamma_i$, $\kappa_i$ is chosen such that the pressure and its first derivative are continuous over the change of the polytropic index and that the lower (higher) edge of the first (last) monotrope matches smoothly to the CET (pQCD) EoS. Also, it is required that the speed of sound stays always sublulminal, $c_s^2 < 1$.

Figure \ref{fig5} (left) demonstrates in practice, how the procedure explained above constrains the EoS at intermediate densities. The figure shows an interpolation constructed from two monotropes of different polytropic indices (dashed line). The nuclear EoS is denoted by the small red segment on the left, whereas the range of possible perturbative EoSs allowed by scale variation is depicted by the orange band on the right. A single realization of the pQCD EoS with a particular choice of the renormalization scale is finally denoted by a dark green line labeled by ``$X=1.2$''. In order for the interpolation to smoothly reach the pQCD EoS at high densities, the first monotrope can not be too soft (cf.~the line with $\gamma_1 = 2.0$); otherwise the pressure irreparably overshoots the pQCD pressure. On the other hand, the pQCD EoS itself is relatively soft with $c_s^2 \lesssim 1/3$, and in order for the last monotrope to smoothly reach pQCD, the latter cannot be too hard (cf.~the line with $\gamma_2 = 4.0$). This gradual softening of the EoS restricts the values of $\gamma_i$ that fulfil both the low and high density constraints. Indeed, fig.~\ref{fig5} (right) displays all the possible polytropic solutions consisting of 2 or 3 monotropes and fulfilling the low and high density constraints, and has been obtained by varying the polytropic indices as well as the chemical potentials where the indices change. It is noteworthy that adding an additional monotrope does not significantly widen the envelope of the set of allowed EoSs, so that the systematic error caused by choosing only 2 monotropes is numerically small.

If there is a first order transition in the intermediate density region, it will appear as a discontinuity of the first derivative of the pressure as a function of $\mu_B$, proportional to the latent heat of the transition when passing from one monotrope to the next. Thermodynamical consistency requires that the discontinuity is positive, such that the EoS is softer (smaller $\gamma_i$) in the high density monotrope than if we have no transition at all. This, combined with the requirement of matching with the pQCD EoS significantly constrains the possible EoSs even with the first order transition. Indeed, assuming a first order transition can be seen to not lead to new solutions outside of the envelope shown in fig.~\ref{fig5}.

The effect of the constraint at high density is perhaps most clearly seen by comparing the interpolated EoS to an extrapolation not taking into account the pQCD constraint. This is shown in fig.~\ref{fig6}, where the cyan and blue bands (including the area of the green band) correspond to the interpolation with two and three monotropes, respectively. The band denoted HLPS is the extrapolation of Hebeler et al.~\cite{Hebeler:2013nza}, who in addition required that the EoS is able to support a 2$M_{\rm sol}$ star and imposed additional \textit{ad hoc} constraints on the $\gamma_i$'s  ($1< \gamma_1 < 4.5$, $0.5 < \gamma_2 < 8.5$).  The additional constraints on the $\gamma_i$'s imposed by Hebeler et al.~are roughly compatible with the constraints determined from pQCD ($2.23 < \gamma_1 < 9.2$, $1.0<\gamma_2 <1.5$), though the pQCD constraint additionally excludes EoSs that are too soft at small densities, so that $\gamma_1 > 2.23$. The pQCD constraint can thus be viewed as providing an a posteriori justification for the choices made in \cite{Hebeler:2013nza}.

Finally, imposing the additional condition on our EoSs that the they must be able to support a 2$M_{\rm sol}$ star leads to the green band of fig.~\ref{fig6}. Summarizing our finding here, using the constraints at low and at high densities, the EoS is known within $\pm 30\%$ at all densities, irrespective of the amount of quark matter that is present in the cores of neutron stars or whether there is a physical phase transition between  the quark matter and nuclear matter phases.

\section{Implications on neutron star properties}

\begin{figure*}[t]
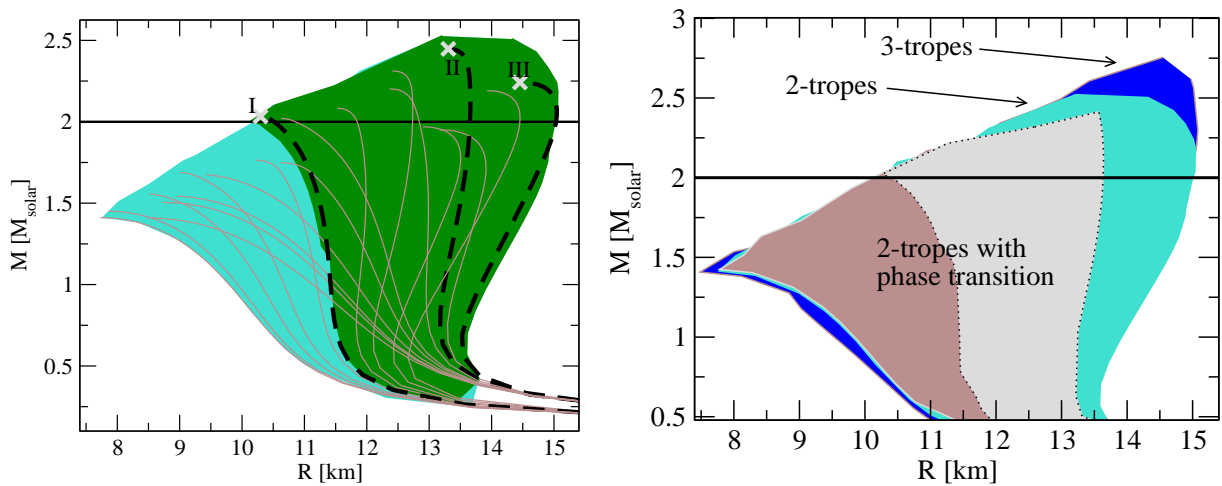

\vspace{0.4cm}
\hspace{0.5cm}\resizebox{0.42\textwidth}{!}{%
\includegraphics{MRv2.eps}}
\hspace{0.1cm}
\resizebox{0.45\textwidth}{!}{%
\includegraphics{bitropes-vs-tritropes-MRv2.eps}}
\caption{Left: Two $M-R$ clouds composed of the EoSs of the same colors displayed in fig.~\ref{fig6}. Right: The $M$-$R$ clouds corresponding to tritropic and first-order phase transition EoSs, shown together with our original result displayed on the left.}
\label{fig7}       
\end{figure*}

In addition to severely constraining the EoS of cold strongly interacting matter, the calculations explained above have implications on the macroscopic properties of neutron stars. Besides the mass-radius relation, one can investigate the internal structure of stars in terms of their energy density distribution, the effects of allowing for a first order transition to quark matter, as well as the dependence of the amount of quark matter inside a star on the latent heat of the transition.

Fig.~\ref{fig7} (left) exhibits two clouds of $M-R$ curves corresponding to all of our generated EoSs. The maximal masses of the stars are seen to fall inside the interval $M_{\rm max}\in [1.4, 2.5]\,M_{\odot}$, while their radii lie in the range $R\in [8,14]$ km. At the same time, the maximal chemical potentials encountered at the center of the star satisfy $\mu_{\rm center} \in [1.33,1.84]$ GeV, corresponding to maximal central densities of $n\in[3.7,14.3]\, n_0$. This falls right in the middle of the interval between the nucleonic and pQCD regions, where the EoS is equally constrained by its low and high density limits. 

As discussed previously, the stellar matter EoS can be further constrained by demanding that it is able to support the observed two solar mass stars. This requirement produces the dark green area in the figure, corresponding to the band of the same color in fig.~\ref{fig6}. For these EoSs, the maximal chemical potentials are bound from above by $\mu_{\rm center}<1.77$ GeV, and the central densities by $n<8.0\, n_0$. From fig.~\ref{fig7}, one can in addition read that for $1.4M_\odot$ neutron stars, our allowed radii range between $11$ and $14.5$ km, while for $2M_\odot$ pulsars, $R\in [10,15]\,$km. It is also worth noting that within the bitrope approach, we find no configurations with masses above $2.5M_\odot$. 

In fig.~\ref{fig7} (right), we add the effects from  tritropic corrections and the presence of a possible first-order phase transition in the EoS for comparison. Both effects are seen to introduce only minor corrections to the earlier results. For a more complete analysis of the case of a first-order phase transition, we refer the reader to \cite{Alford:2013aca}, where the authors also invvestigate the case of twin star configurations, which we have completely omitted in our study.

\begin{figure*}[t]
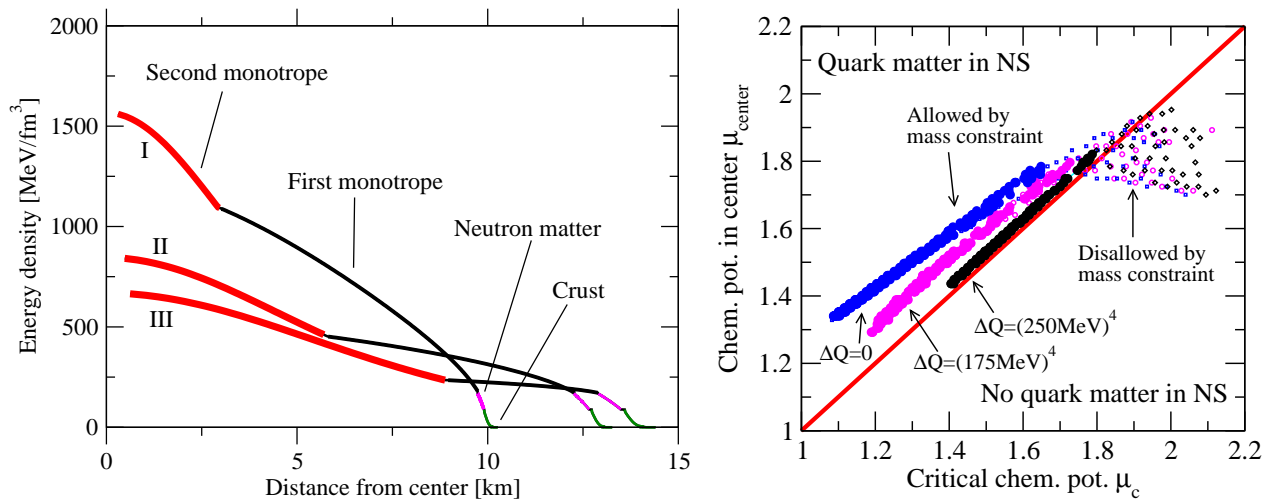

\vspace{0.9cm}
\hspace{0.5cm}\resizebox{0.50\textwidth}{!}{%
\includegraphics{inner_structure.eps}}
\hspace{0.1cm}
\resizebox{0.4\textwidth}{!}{%
\includegraphics{mumaxVsmux_v2.eps}}
\caption{Left: The internal structure of the maximally massive stars corresponding to three representative EoSs I-III (see  \cite{Kurkela:2014vha} for details). Right: The maximal chemical potential reached at the center of a maximum mass star as a function of the critical chemical potential. See the main text for details.}
\label{fig-quark matter}       
\end{figure*}

In Fig. \ref{fig-quark matter} (left), we finally show the inner structure of three maximally massive stars corresponding to the cases I-III of fig.~\ref{fig7}. The energy densities are all continuous due to the smoothness of the matching procedure (no first-order phase transition), while one can witness the softening of the EoS when approaching the perturbative densities as the faster growth of the energy density near the center of the star. Fig.~\ref{fig-quark matter} (right) on the other hand displays the maximal chemical potential reached at the center of a maximum mass star as a function of the critical chemical potential. As a test on the effects of a possible first-order transition to quark matter, we consider several values of the parameter $\Delta Q$, standing for the strength (latent heat) of the phase transition: $\Delta Q=0$ (blue), $\Delta Q=(175\rm{MeV})^4$ (magenta), and $\Delta Q=(250\rm{MeV})^4$ (black). The open points correspond to EoSs that cannot support a $M=2M_{\odot}$ star, while the solid points are allowed by the mass constraint and the diagonal line separates the region that allows for the presence of deconfined quark matter in the core of the compact star from the one where no quark matter phase is realized.

\section{Final remarks}

One of the main challenges in the investigation of the microscopic and macroscopic properties of neutron stars is how to account for the Equation of State of strongly interacting matter in the relevant density interval, covering possibly both confined and deconfined phases. First-principle approaches to the problem are unfortunately scarcely available, given the intrinsic difficulty of lattice QCD  at nonzero baryon density. The most common alternative is to resort to one of the multitude of different models that try to capture some features of the fundamental theory. Their use is, however, typically accompanied by the introduction of systematic errors that are often hard to quantify. 

As a viable alternative to models, there are at least two limits of phenomenological relevance in which QCD admits a first-principle  approach. One of these is the limit of high densities, where pQCD provides robust results that furthermore come with a built-in measure of their inherent uncertainties. Up to now, these results have, however, not been fully taken advantage of by the neutron star community, even though several attempts in this direction have been made over the years \cite{Kurkela:2014vha,Kurkela:2009gj,Fraga:2013qra,Fraga:2001id,Fraga:2001xc,Alford:2004pf,Fraga:2004gz,Kurkela:2010yk}. 

In the paper at hand, we  have discussed, how the EoS for cold nuclear matter can be constrained by using first principles results from  two opposite limits in baryon density and combining these insights together with observational constraints on the maximum mass of neutron stars. As discussed in detail in \cite{Kurkela:2014vha}, the strong constraints that emerge from this are remarkably independent of the existence of deconfined quark matter in the cores of neutron stars. This demonstrates that quark matter physics is relevant for the equation of state for cold nuclear matter even at densities below the critical density for the hadron-quark transition.

\vspace{0.5cm}
\noindent
{\it Acknowledgments:} 
The authors are grateful to J. Schaffner-Bielich for fruitful discussions and collaboration. ESF is supported by CAPES, CNPq and FAPERJ, while AV is supported by the Academy of Finland, grant nr. 266185.


\end{document}